\documentclass[preprint2]{aastex}

\shorttitle{A Tunable Echelle Imager}
\shortauthors{I.~K.~Baldry and J.~Bland-Hawthorn}

\newcommand{\micr}{\,$\mu$m}

\begin{document}

\title{A Tunable Echelle Imager}
\author{I.~K.~Baldry and J.~Bland-Hawthorn}
\affil{Anglo-Australian Observatory}
\affil{P.O.\,Box 296, Epping, NSW 1710, Australia}
\email{baldry@aaoepp.aao.gov.au, jbh@aaoepp.aao.gov.au}
\date{Received 2000 March 14; accepted 2000 April 19}

\begin{abstract}
We describe and evaluate a new instrument design called a 
Tunable Echelle Imager (TEI). 
In this instrument, the output from an imaging Fabry-Perot interferometer 
is cross-dispersed by a grism in one direction and dispersed by 
an echelle grating in the perpendicular direction. 
This forms a mosaic of different narrow-band images of the same field 
on a detector. 
It offers a distinct wavelength multiplex advantage over a traditional 
imaging Fabry-Perot device. 

Potential applications of the TEI include spectrophotometric imaging 
and OH-suppressed imaging by rejection. 
\end{abstract}


\section{Introduction}
\label{sec:intro}

\begin{figure}
 \plotone{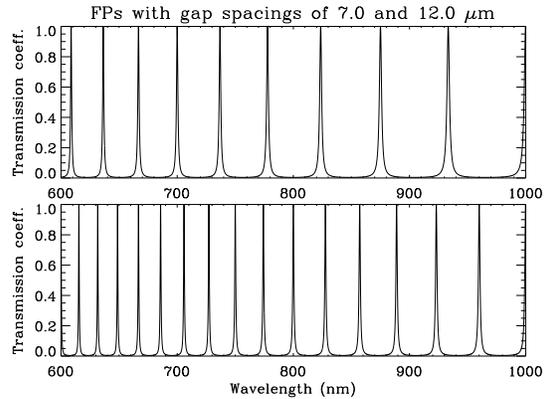}
 \caption{Transmission profiles for Fabry-Perot filters with two different
   gap spacings.  The upper plot shows orders from 14 on the right to
   23 on the left; the lower plot shows orders 24 to~40.}
 \label{fig:gap-sp}
\end{figure}

One of the most versatile instruments for astronomical narrow-band imaging 
is the gap-scanning Fabry-Perot (FP) etalon. 
Light passing through a FP produces a series of wavelength passbands 
called orders (examples of the transmission are shown in 
Figure~\ref{fig:gap-sp}). 
The positions of the passbands can be moved by varying the separation
of the cavity surfaces \citep[e.g.,][]{ARRH81,Ath95,BJ98pasa}.
However, in order to image a field with a FP, broad- or
intermediate-band filters are usually used to block out the light from
all the passbands except the one of interest.  An alternative method
is to use a dispersive element to separate the images formed by
different FP orders. 
Recently this principle has been used in the PYTHEAS instrument 
\citep{LBC95} and the GraF instrument 
(\citeauthor{CL95} \citeyear{CL95}; \citeauthor{CLL99}
\citeyear{CLL99}, \citeyear{CLR99}).
In this paper, we consider the use of a cross-dispersed echelle
grating\footnote{We use the term `echelle' for any multi-order
  diffraction grating used above about the fifth order, rather than
  distinguishing between an `echellette' and an `echelle'
  grating.} crossed with a gap-scanning FP called a 
Tunable Echelle Imager (TEI).
This forms a mosaic of narrow-band images with different wavelengths 
on a detector, and by scanning the plate separation of the FP, 
a complete data cube can be obtained (2D image plus 1D spectral). 

\subsection{The basic physics}

The wavelength positions of the FP orders are described by the
equation for constructive interference:\footnote{Note that
  Equation~\ref{eqn:orders-fp} does not include the effect of phase
  changes on reflection \citep{ARRH81,JB98}.  Such phase changes only
  have a significant impact for low-order etalons which are not
  considered here.}
\begin{equation}
 m_{\rm fp} \lambda = 2 \mu d \cos \theta \: ,
 \label{eqn:orders-fp}
\end{equation}
where $m_{\rm fp}$ is the FP order, $\mu$ is the refractive index of 
the medium in the cavity gap, $d$ is the plate separation (or gap spacing) 
and $\theta$ is the angle of incidence. 

The bandwidth of each FP order is dependent on the reflectivity and 
defects of the coated surfaces (non-parallelism). 
The ratio between inter-order wavelength spacing ($\Delta \lambda$) 
and the FWHM bandwidth ($\delta \lambda$) is called the effective finesse: 
\begin{equation}
 N = \frac{\Delta \lambda}{\delta \lambda} \: .
 \label{eqn:finesse}
\end{equation}
The finesse is also related to 
the contrast between the peak transmission of an order and the 
minimum transmission between orders. 
For contrast values greater than 100, 250, 500 or 1000, the 
finesse needs to be greater than 16, 25, 35 or 50, respectively. 
At high finesse, the efficiency can drop significantly due to defects 
\citep{ARRH81}. 
A finesse between 20 and 40 is optimal for a high-throughput imaging FP
\citep{Bla95}. 

The echelle grating (EG) disperses the FP orders using multi-order 
diffraction. The diffracted angle ($\beta$) for gratings 
that are not immersed between prisms is determined from the 
grating equation: 
\begin{equation}
 m_{\rm eg} \lambda = \mu \Lambda (\sin \alpha + \sin \beta) \: ,
 \label{eqn:orders-eg}
\end{equation}
where $m_{\rm eg}$ is the EG order, $\mu$ is the refractive 
index of air, $\Lambda$ is the grating period and $\alpha$ is the 
incidence angle in air. 

With an echelle grating, several wavelengths corresponding to different 
EG orders are diffracted at or near the same angle. 
For this reason, a low-dispersion element is needed to separate the 
EG orders. 
This could be another grating, one or more prisms, or a grism. 

\section{TEI setup}
\label{sec:setup}

A TEI system is suitable for use at Cassegrain focus 
with a small collimated beam, 
preferably less than 100\,mm in diameter 
for cheaper and better-quality etalons. 

\begin{figure}
 \plotone{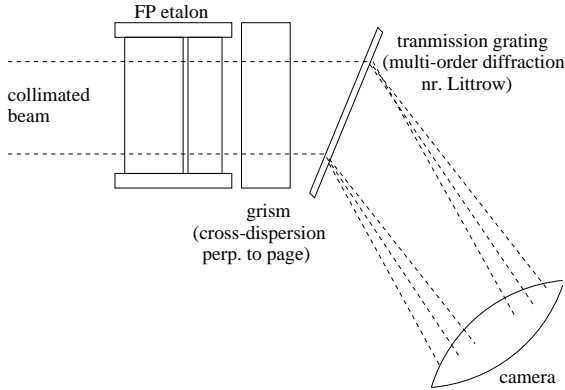}
 \caption{TEI with a transmission echelle grating. 
   The grism disperses the FP orders perpendicular to the page. The
   echelle grating disperses the light near Littrow thus preserving
   the aspect ratio of the field on the detector. The many orders of
   the echelle allow a mosaic of images to be formed.}
 \label{fig:echgraf-trans}
\end{figure}

\begin{figure}
 \plotone{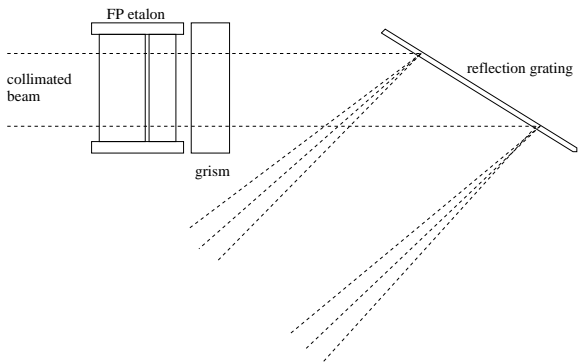}
 \caption{TEI with a reflection echelle grating (blaze to collimator).
   In this case, the echelle grating expands the beam.}
 \label{fig:echgraf-reflec1}
\end{figure}

\begin{figure}
 \plotone{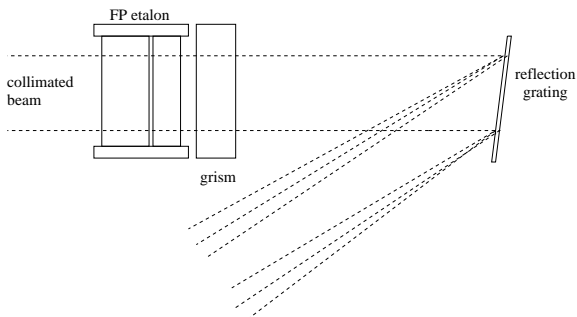}
 \caption{TEI with a reflection echelle grating (blaze to camera).
   In this case, the echelle grating reduces the size of the beam.}
 \label{fig:echgraf-reflec2}
\end{figure}

An aperture is placed in the telescope focal plane in order to prevent 
overlap of images at the detector. 
The TEI is placed in the collimated beam. 
Different setups are shown in Figures~\ref{fig:echgraf-trans}, 
\ref{fig:echgraf-reflec1} and~\ref{fig:echgraf-reflec2}.

The first optical element is the Fabry-Perot (FP) filter which 
produces a series of narrow passbands or orders. 
Alternatively, 
it is possible to put one or both of the dispersing elements before the FP. 
The advantage in placing the FP first is that all the light 
from a given point source passes through the FP at the same angle ($\theta$). 
This allows for a smaller etalon and the 
variations in central wavelength across a field are the same for all orders 
(consider the $\cos \theta$ factor in Equation~\ref{eqn:orders-fp}). 
If the variation in $\theta$ is small enough, the image of each field will be 
effectively monochromatic. 
For monochromaticity, it is preferable that the centre of the field is 
on-axis with the FP, i.e., to minimize the variation in $\cos \theta$.

\begin{figure}
 \plotone{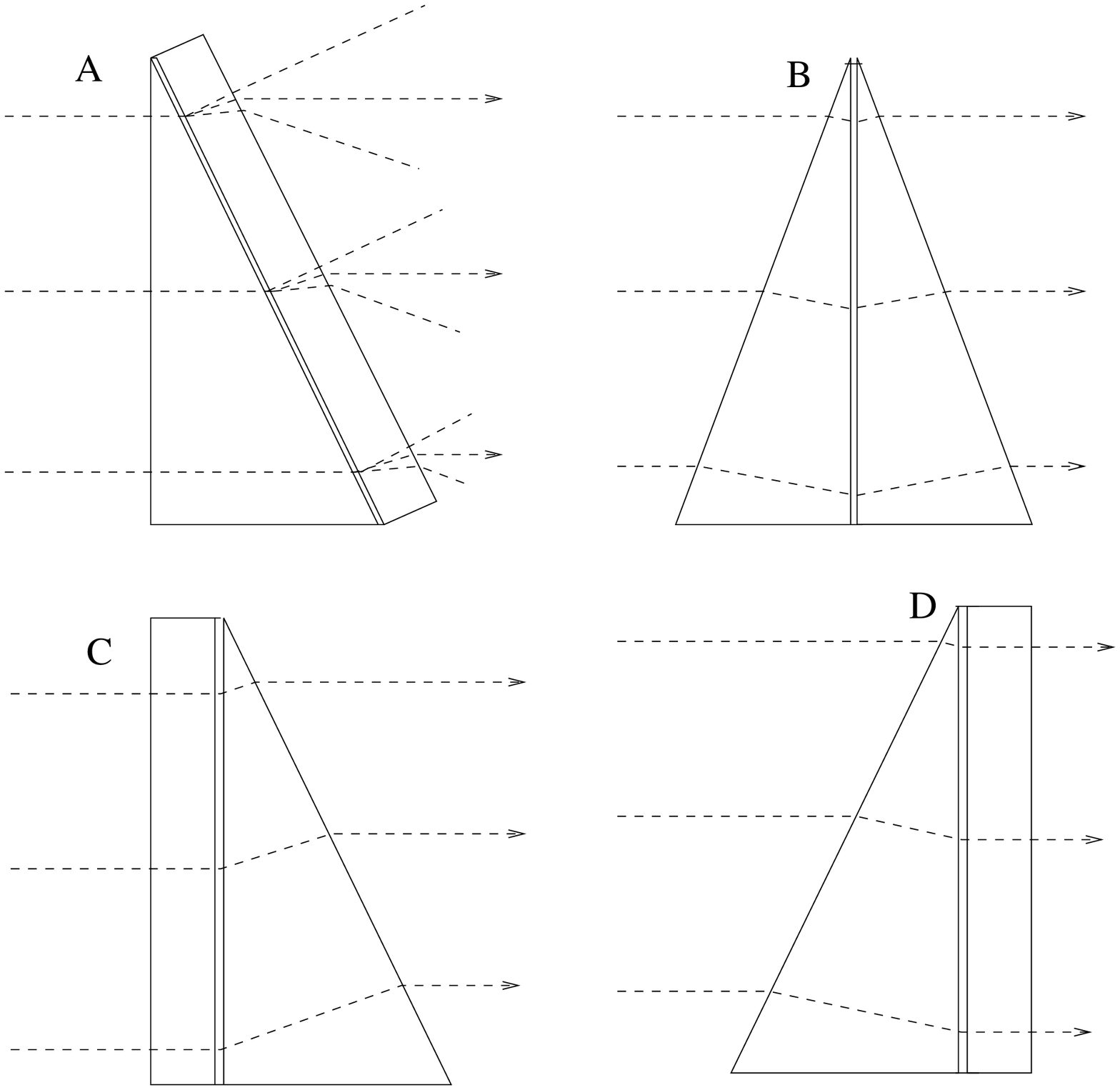}
 \caption{Examples of grism formats. 
   With A- and B-type formats, the beam remains the same size for zero
   deviation light. The C-type format reduces the size of the beam
   while the D-type format expands the beam slightly.}
 \label{fig:grisms}
\end{figure}

The second optical element is a grism which separates the FP orders 
with low-dispersion first-order diffraction. 
In principal, a grating without a prism could be used. 
The advantage of using a grism is that it is more likely to be compatible 
with existing astronomical optical instruments 
if the beam is not deviated significantly at this point. 
Some grism formats are shown in Figure~\ref{fig:grisms}. 
In some cases, it may be necessary to include a colored-glass filter 
to block second- and higher-order diffraction by the grism. 
This could be placed before the FP element to minimize scattered 
light in the TEI. 

The third optical element disperses the light in the perpendicular
direction (to the grism) with high-dispersion multi-order diffraction.
The dispersion is too high to correct to zero beam deviation by using
a prism.  The setup in Figure~\ref{fig:echgraf-trans} uses a
transmission grating with Littrow diffraction. The advantage of this
is that the beam size remains the same after diffraction and,
therefore, the pixel scale on the detector is the same as for direct
imaging.  The setup in Figure~\ref{fig:echgraf-reflec1} uses a
reflection grating.  Higher spectral resolution (or a larger field for
a given resolution) can be achieved with this setup but there is an
anamorphic increase in beam size.  An alternative reflection grating
setup is shown in Figure~\ref{fig:echgraf-reflec2} with 
an anamorphic decrease in beam size.  In this case, 
it may be advantageous to place the grism after the reflection grating.

The use of an articulated camera\footnote{For example, articulated
  cameras are planned for the ATLAS \citep{RTB00} and OSIRIS
  \citep{CAG00} spectrographs.} would allow a choice of the beam
deviation that is most suitable for the experiment.  Especially in the
case of the transmission echelle grating, the ability to control the
observed angle of diffraction allows significant control over the
dispersion and anamorphic factor of the instrument.  The recent advent
of volume-phase holographic (VPH) technology offers another advantage
for the completely transmissive version of the TEI, in that, the grism
and echelle grating could be combined into one optical element for
higher efficiency (a multiplexed grating; \citeauthor{BWAC00}
\citeyear{BWAC00}).

\subsection{Parameters}

There are a number of different parameters that define the properties of 
the instrument. 
The principal parameters that we will consider are: 
\begin{itemize}
\item for the telescope: 
  \begin{itemize}
  \item the field-of-view size, 
  \item the beam reduction factor (collimated-beam diameter /
    telescope diameter $= f_{\rm coll} / f_{\rm tel}$), which
    determines the scaling between angles in the collimated beam and
    on the sky;
  \end{itemize}
\item for the Fabry-Perot filter:
  \begin{itemize}
  \item the separation between the highly reflective surfaces of the
    FP ($d$), which determines the central wavelengths of the orders
    and the separations between the orders,
  \item the finesse ($N$), which determines the resolution and
    contrast for a given order;
  \end{itemize}
\item for the echelle grating:
  \begin{itemize}
  \item the line density of the grating ($1/\Lambda$),
  \item the incident and diffracted angles ($\alpha$ and $\beta$).
  \end{itemize}
\item for the grism:
  \begin{itemize}
  \item the line density of the grating,
  \item the angle and the refractive index of the prism(s),
  \item the incident and diffracted angles;
  \end{itemize}
\end{itemize}

In the next section (\ref{sec:examples}), we give some examples of 
various setup parameters. 

\subsection{Examples}
\label{sec:examples}

To determine the type of images that can be obtained with a TEI, 
we shall consider a system that can image a 12\degr\,x\,12\degr\ angular 
spread in the collimated beam. This corresponds to a 6\,x\,6\,arcminute 
field of view for a telescope with a beam reduction factor of 1/120
(e.g., a 10-m telescope with an 83-mm collimated beam), 
or a 12\,x\,12\,arcminute view with a beam reduction factor of 1/60 
(e.g., a 4-m telescope with a 67-mm beam). 
For the 6\,x\,6\,arcminute field, the image sampling corresponds to 
0.18\,arcsec per pixel for a 2K\,x\,2K detector. 

Examples of various setup parameters are shown in Table~\ref{tab:setup-para}. 
\begin{table*}
 \caption{Examples of various setup parameters for a TEI}
 \label{tab:setup-para}
 \vspace{0.3cm}
 \begin{tabular}{ccccccccccc} \hline \hline 
 Aperture &\multicolumn{3}{l}{Fabry-Perot parameters .....} 
 &\multicolumn{6}{l}{Echelle-grating parameters .....} &Fig. \\ \hline
 Field size$^a$ &Plate sep. &Res.$^b$~at &No.$^c$~of
 &Type &Line~d. &Inc. &Diff.$^d$ &Anamor. &No.$^e$~of & \\
 (arcsec) &($\mu$m) &600\,nm &orders
 & &(mm$^{-1}$) &angle &angle &factor &orders & \\ 
 \hline
 48\,x\,38&~~12&~~1200&~~17&trans. &145&25\degr&25\degr&1.00&~5&\ref{fig:im1}\\
 39\,x\,23&~~27&~~2700&~~37&trans. &145&32\degr&32\degr&1.00&~6&\ref{fig:im2}\\
 39\,x\,14&~~45&~~4500&~~61&trans. &145&32\degr&32\degr&1.00&~6&---\\
 19\,x\,11&~~80&~~8000&~107&trans. &~90&40\degr&40\degr&1.00&11&\ref{fig:im3}\\
 20\,x\,~9&~~80&~~8000&~107&reflec.&~90&25\degr&55\degr&0.63&10&\ref{fig:im4}\\
 20\,x\,14&~~80&~~8000&~107&reflec.&~90&55\degr&25\degr&1.58&10&\ref{fig:im5}\\
 12\,x\,~6&~190&~19000&~254&reflec.&~70&43\degr&63\degr&0.62&16&\ref{fig:im6}\\
 12\,x\,~6&~310&~31000&~414&reflec.&~70&63\degr&43\degr&1.61&16&---\\
 15\,x\,~4&~900&~90000&1201&reflec.&100&75\degr&55\degr&2.22&13&---\\
 \hline
 \end{tabular} \vspace{0.2cm} \\
 Note: all the examples use a grism (type-A format, 410\,lines/mm,
 29\degr-prism, $n_{\rm prism}=1.67$) which gives zero deviation at a
 wavelength of 800\,nm with first-order diffraction. For the
 reflection echelle gratings, the angle between camera and collimator
 is 20\degr\ or 30\degr. \\ 
 \llap{$^a$}The field size chosen to match the setup with no overlap
 of imaged FP orders, assuming a beam reduction factor of 1/120 (e.g.,
 a 10-m telescope with an 83-mm collimated beam).  For a beam
 reduction factor of 1/60 (e.g., a 4-m telescope with a 67-mm beam),
 the field dimensions are doubled. \\ 
 \llap{$^b$}The resolving power ($\lambda/\delta \lambda$) at 600\,nm, 
 assuming a finesse of 30, $\theta = 0$ and $\mu = 1$ for the FP. \\
 \llap{$^c$}The number of FP orders between 600 and 1000\,nm. \\
 \llap{$^d$}The diffracted angle required to image on the central axis 
 of the detector (ideally peak efficiency). \\
 \llap{$^e$}The number of useful EG orders for the 
 wavelengths between 600 and 1000\,nm. 
\end{table*}
The fixed grism format disperses the wavelength range 600 to 1000\,nm
across the detector with first-order diffraction.
For most optical systems, a colored filter would be required to block 
wavelengths below 500\,nm (e.g., OG\,550) to avoid contamination 
of the images with second-order diffraction from the grism.
This range was chosen to span roughly an octave in spectral coverage. 
The TEI setup parameters can easily be adapted for other 
wavelength ranges, beam characteristics and detector sizes. 

The plate separation of the FP determines the number of imaged orders
between 600 and 1000\,nm. This combined with the finesse determines
the resolving power of the instrument. If the finesse is constant
across the observed wavelengths then the resolving power will be
highest at the shortest wavelength and be proportional to $1/\lambda$.
This could be altered by increasing the reflectivity as a function of
wavelength.

\begin{figure}
 \plotone{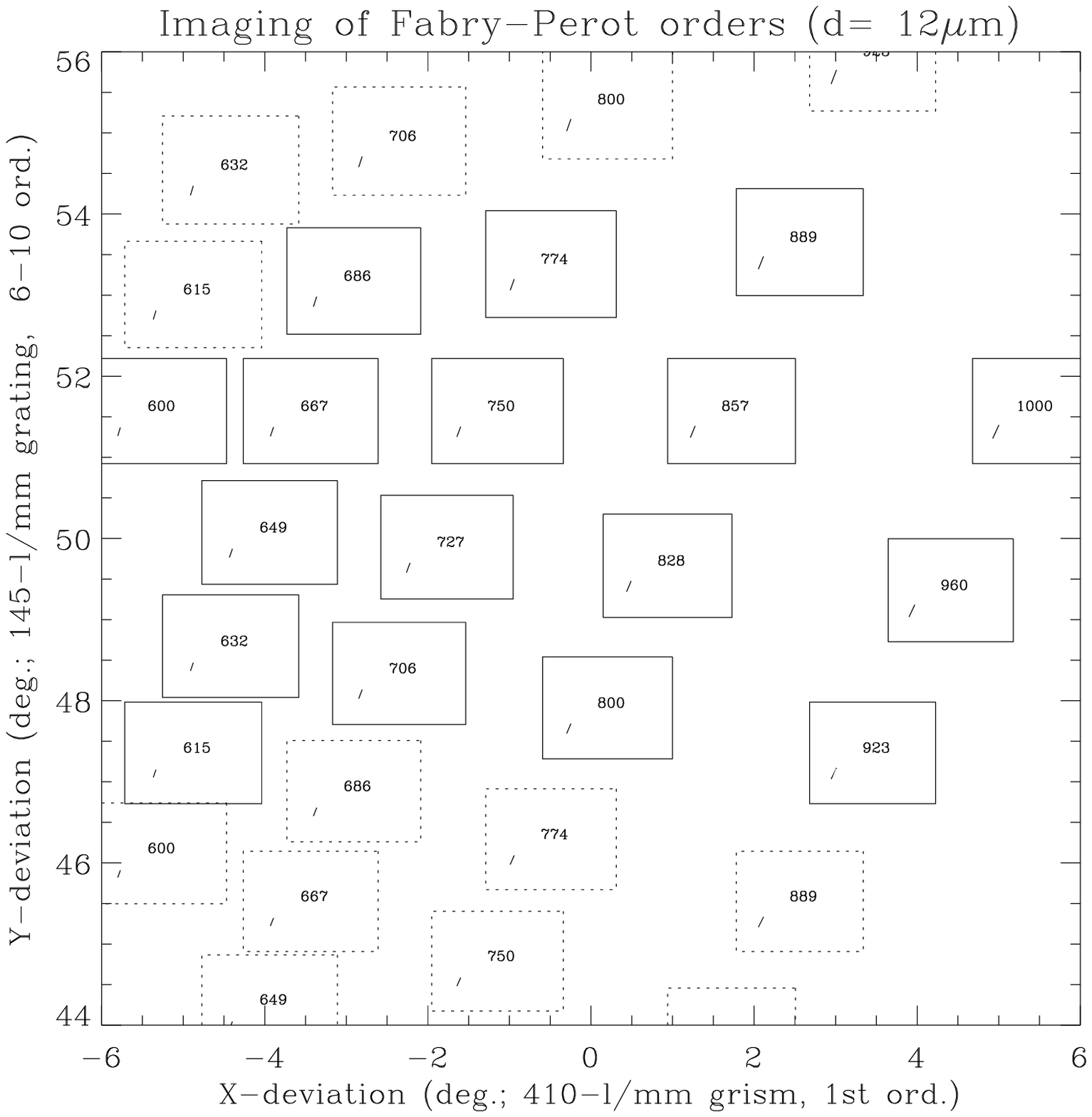}
 \caption{Mosaic of images produced by a TEI. 
   The angles relate to deviations within the collimated beam (not to
   angles on the sky).  The central wavelengths in nanometres of each
   narrow-band image are shown within each square. Images shown with 
   dashed lines are repeated FP orders. The line in the bottom-left 
   corner of each image is the PSF caused by the dispersion.}
 \label{fig:im1}
\end{figure}

\begin{figure}
 \plotone{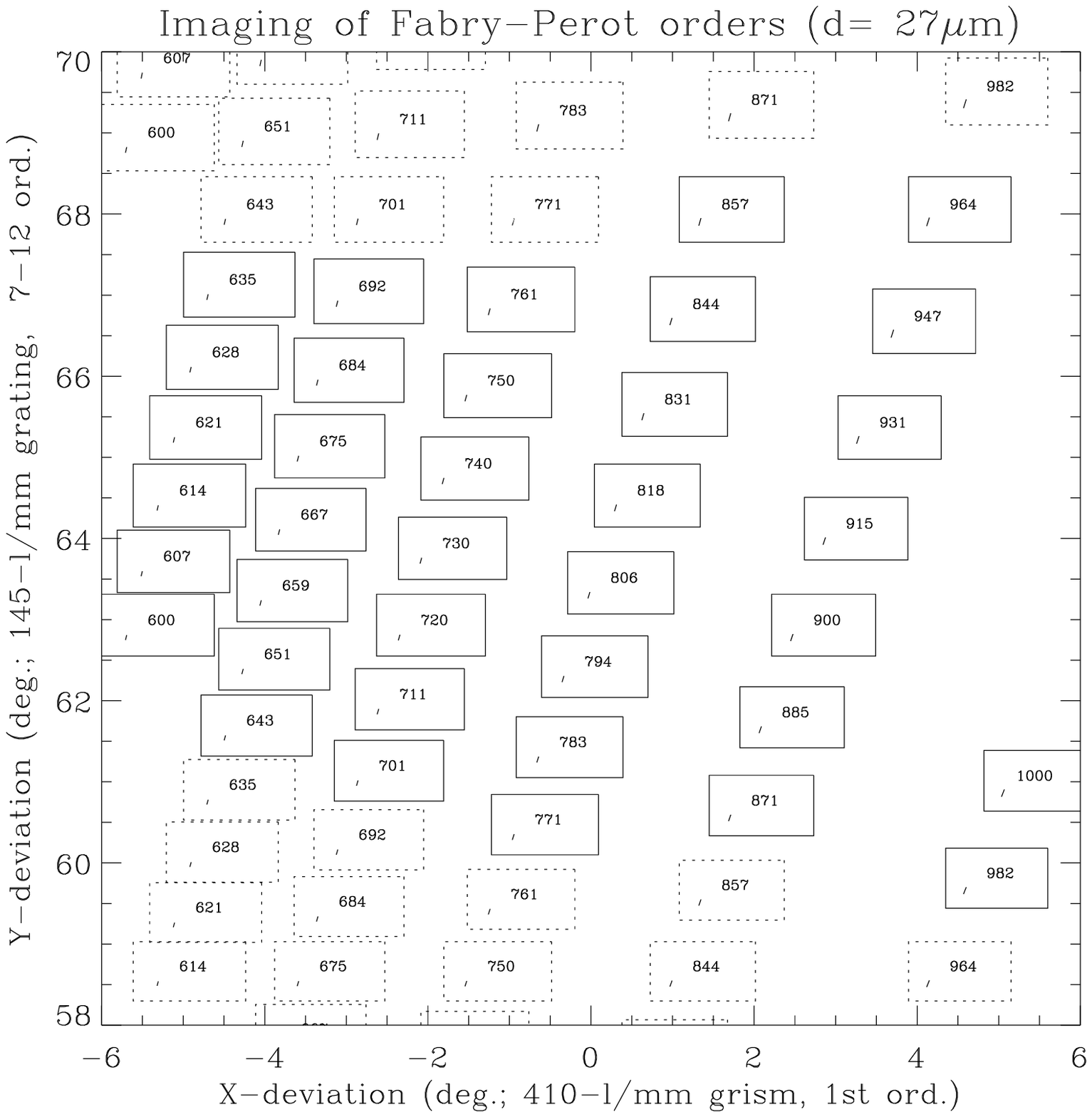}
 \caption{Mosaic of images produced by a TEI.}
 \label{fig:im2}
\end{figure}

\begin{figure}
 \plotone{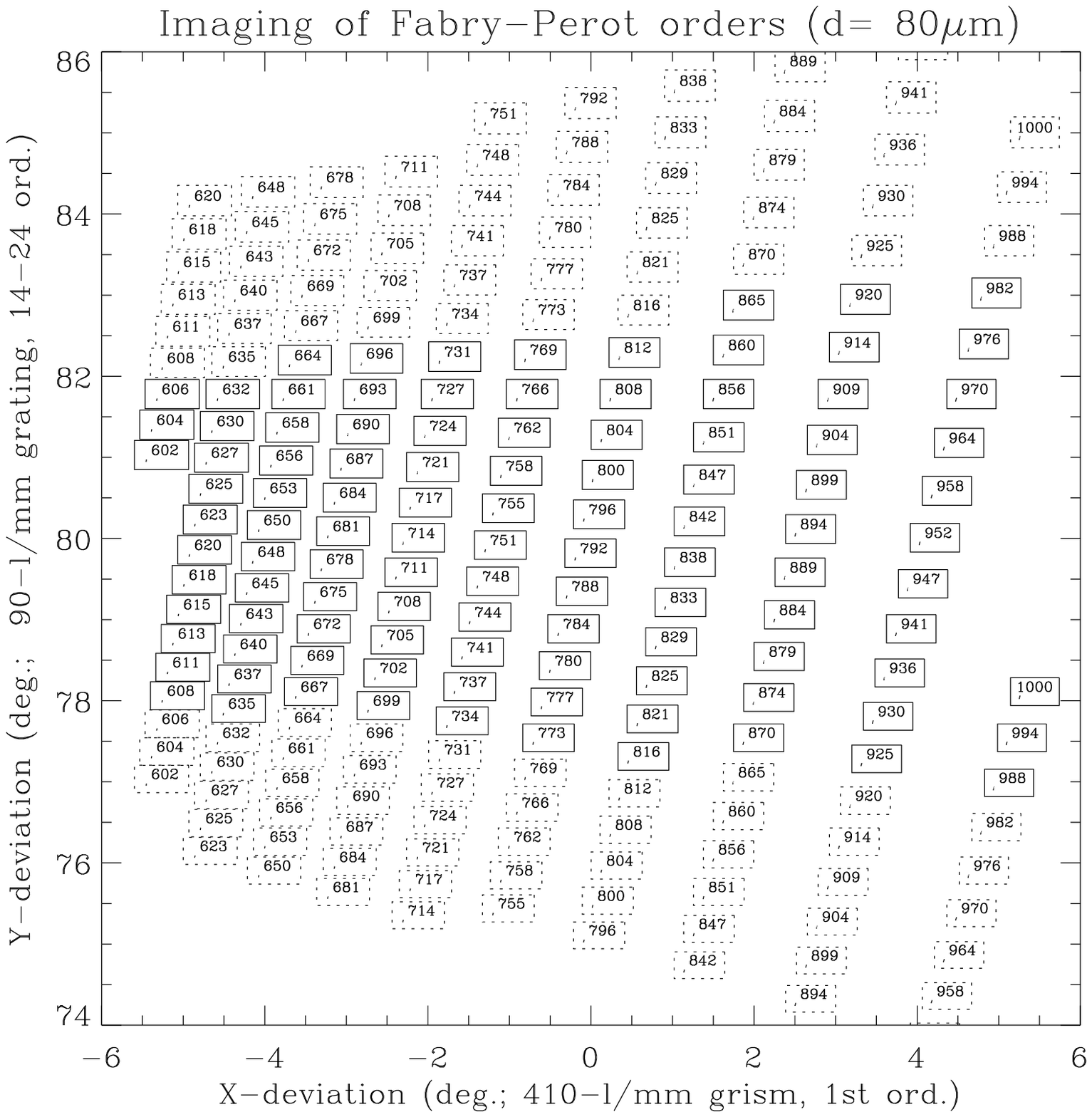}
 \caption{Mosaic of images produced by a TEI.}
 \label{fig:im3}
\end{figure}

\begin{figure}
 \plotone{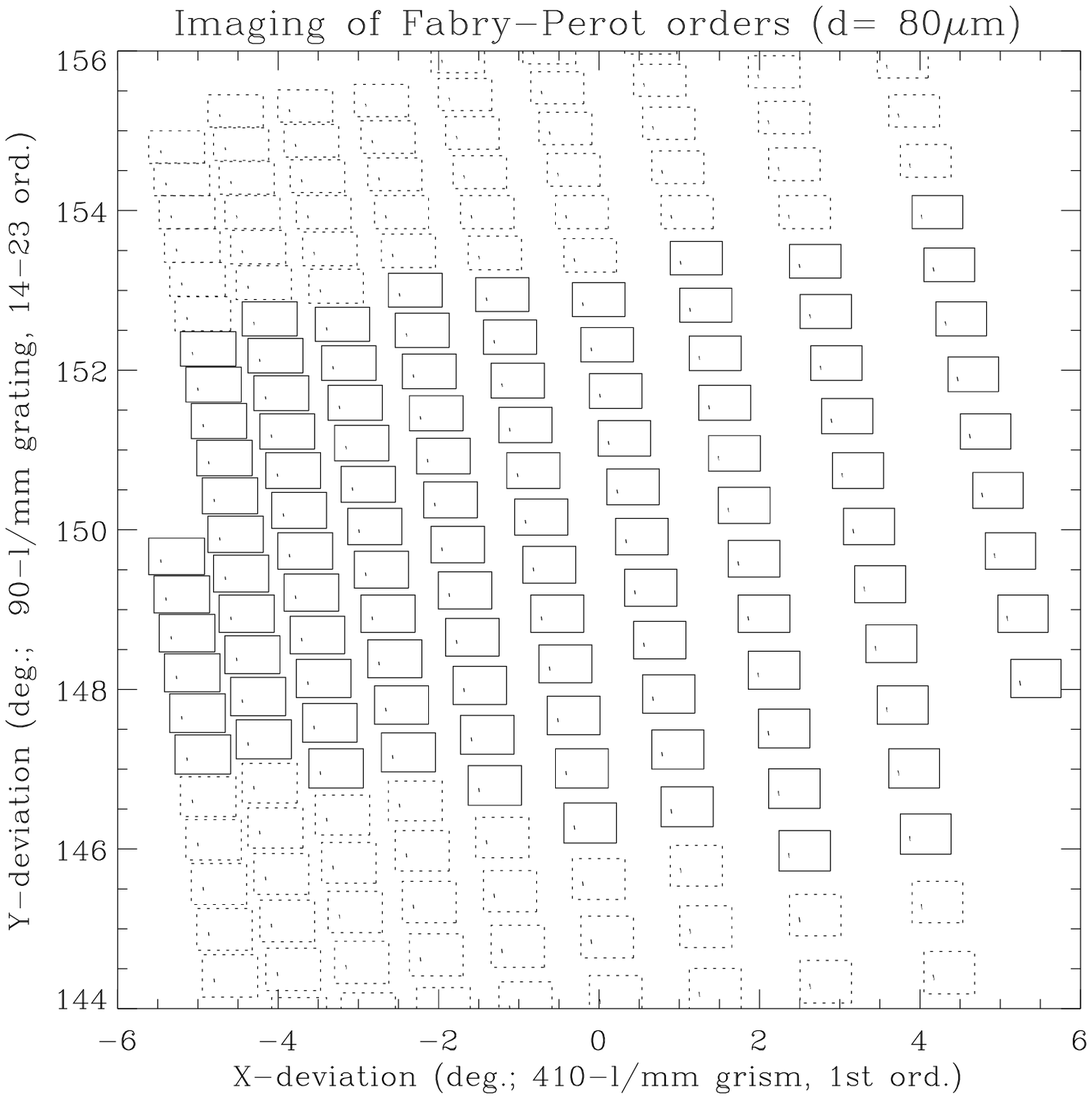}
 \caption{Mosaic of images produced by a TEI.}
 \label{fig:im4}
\end{figure}

\begin{figure}
 \plotone{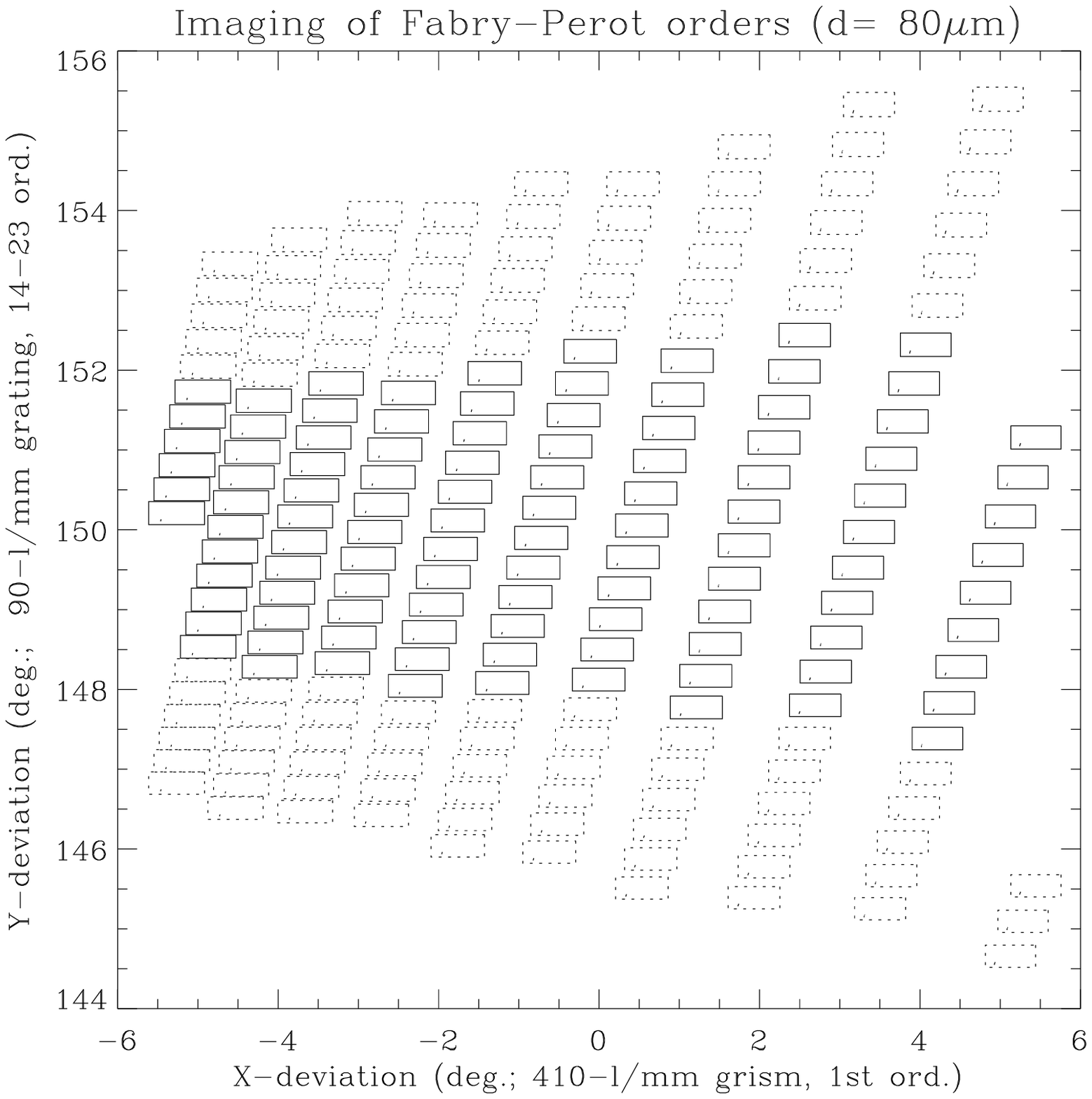}
 \caption{Mosaic of images produced by a TEI.}
 \label{fig:im5}
\end{figure}

\begin{figure}
 \plotone{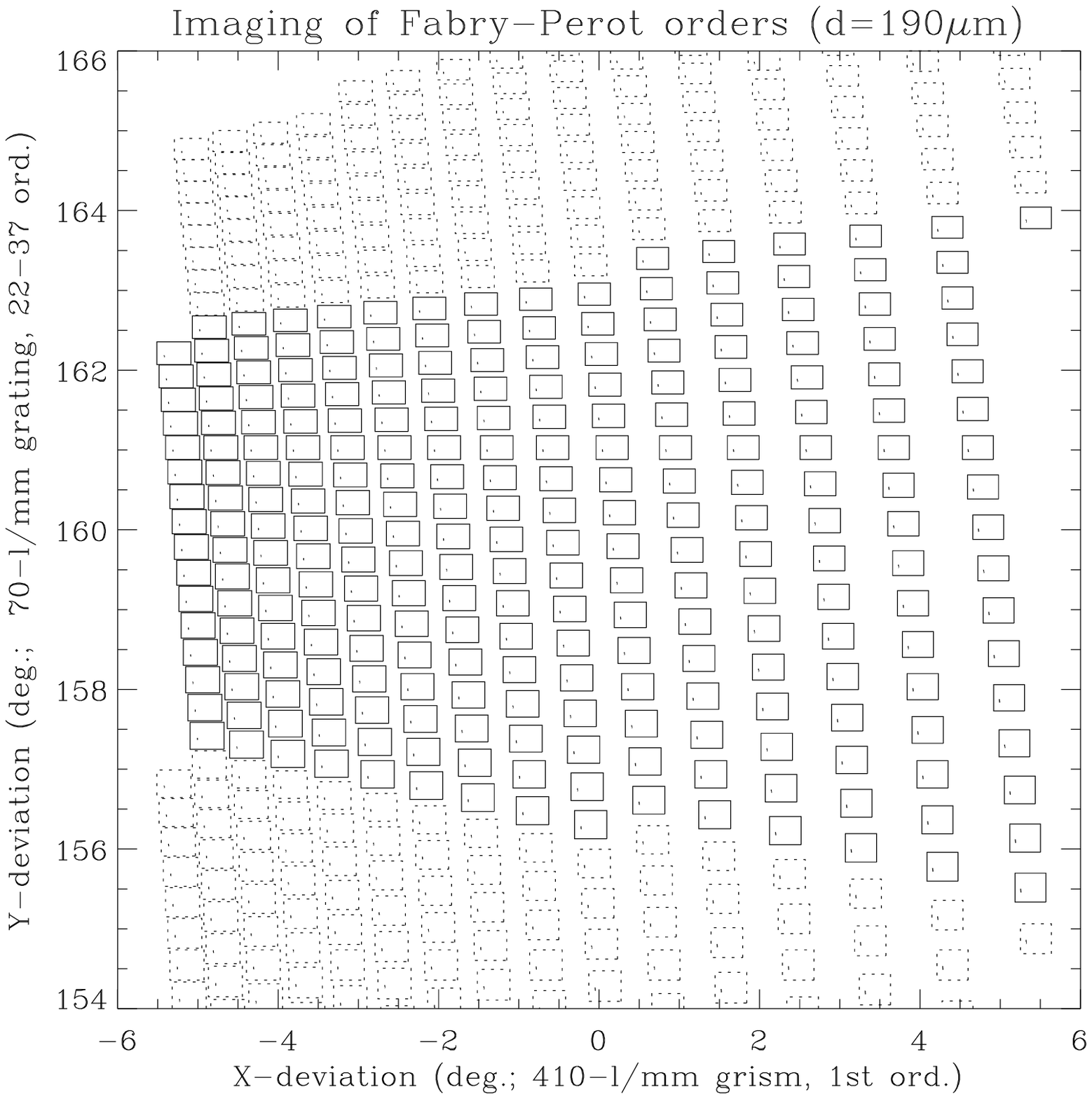}
 \caption{Mosaic of images produced by a TEI.}
 \label{fig:im6}
\end{figure}

The parameters of the echelle grating determine the maximum aperture
size usable at a given resolving power (without overlap of FP orders). 
Higher dispersion gratings and/or increased anamorphic factor allow for 
wider apertures or a larger resolving power. 
The maximum length of the aperture is determined by
the separation of the EG orders imaged onto the detector.
Simulated images of FP orders with various EG orders
and dispersions (Table~\ref{tab:setup-para}) are shown in
Figures~\ref{fig:im1}--\ref{fig:im6}.

\section{Discussion}

\subsection{Observing and data reduction}

To fully sample the spectral dimension at the resolving power of the
TEI setup, it will be necessary to take a number of exposures (at
different FP gap spacings) equal to or greater than the finesse of the
FP.  The reason for a greater number of exposures is because the
requirements for sampling at the shortest wavelength may be different
from the longest wavelength.  In many cases, there will be double
sampling of some wavelengths near the shortest wavelength, these could
be used as a self-calibration check in reconstructing the spectral
dimension.\footnote{\cite{BT89} noted that it was easier to
  reconstruct the $\lambda$ dimension than an image dimension (e.g.,
  as with a scanning slit or IFU).} Spectrophotometric measurements
could be obtained in photometric conditions by calibrating with a
standard star.

The TEI can mimic a true 3D spectrograph by scanning, as described above. 
Some applications may not need scanning or only partial scanning 
(a few exposures) if the sampling of the spectral dimension is adequate.
For example, multiple narrow-band imaging could be used to selectively 
combine images in order to suppress sky-emission lines, 
thus mimicking an OH-suppression instrument. 

Wavelength calibration of a TEI would be made using 
arc lamps to define the wavelength for each order 
and gap spacing for the etalon. 
The calibration is then uniquely defined for a given gap spacing 
and should be stable during an exposure using capacitance 
micrometry \citep{HRA84}. 
Advantageously, the spectral stability of the TEI is not subject to 
instrument flexure as with a normal Cassegrain spectrograph. 

In some of the example image formats, it appears that optimal use is 
not made of the detector in the sense that there are repeated 
orders (e.g., Figure~\ref{fig:im5}). 
However, if every order is repeated twice on the detector, 
it may increase the efficiency and allow the possibility of 
removing image defects (e.g., hot pixels, cosmic rays, bad columns) 
by comparing the same narrow-band image. 

\subsection{Aperture width and the PSF}

The dispersion of the gratings spreads out the light from a single FP
order, contributing to the point spread function (PSF) of the image.
The FWHM of this one-dimensional contribution is equal to the
separation (on the detector) between adjacent FP orders divided by the
finesse.  The lines in the bottom-left corner of each image in the
Figures~\ref{fig:im1}--\ref{fig:im6} represent this PSF.

If the aperture size is maximized for a particular echelle grating and FP
setup then the FWHM of the dispersive PSF is equal to the width of the 
aperture divided by the finesse, for the lowest wavelength. 
Therefore, if a certain image resolution is required in both spatial dimensions
(the TEI only degrades resolution in one direction), then there is a limit
on the setups to those with certain aperture widths. 
For example, in the case of adaptive optics a resolution of 0.1\,arcsec 
may be required limiting the aperture width to 3\,arcsec (with a finesse 
of 30). Alternatively, if 0.5\,arcsec is satisfactory then an aperture 
width of 15\,arcsec would be suitable. 
If good image resolution is only required in one direction then the 
aperture width can be larger still. 

We note that the elongated PSF is a common occurrence in radio interferometry 
where many software tools exist to treat such data \citep{PSB94}. 
The TEI situation is somewhat better because an instrument rotation 
allows us to sample the field with orthogonal PSF orientations. 

\subsection{Anamorphic factor}

The ratio of the beam sizes after and before diffraction 
by a grating is called the anamorphic factor. 
For a reflection grating or a transmission grating without prisms, 
the anamorphic factor 
\begin{equation}
 A = \frac{\cos \beta}{\cos \alpha} \: ,
 \label{eqn:anamorphic}
\end{equation} 
where $\alpha$ and $\beta$ are the incident and diffracted angles in
air.  The change in beam size is inverse to the change in angular
size (the angle between light rays in the collimated beam).
A grating setup which has an anamorphic factor above unity can
increase the resolution because the angular size of the aperture (in
the dispersion direction) is reduced. However, the increase in
resolution is less than the anamorphic factor because the dispersion
is also reduced (though not by as much).

Figures~\ref{fig:im4} to~\ref{fig:im5} show the effect of varying the 
anamorphic factor with the same resolution and EG orders 
(see Table~\ref{tab:setup-para}). 
Larger apertures can be used with the blaze-to-collimator setup 
(increasing beam size) even though the images appear smaller on the detector
(Fig.~\ref{fig:im5}). 
Using the example of a 2K\,x\,2K CCD and a 6\,x\,6\,arcminute field, 
the pixel sampling on the detector would change 
from 0.18 to 0.28\,arcsec/pixel across the width of the aperture (14\,arcsec).
The change in sampling is not critical in this case because the 
dispersion PSF is about 0.47\,arcsec (FWHM). In quadrature with 
0.5\,arcsec seeing, we find that the sampling is still adequate. 
In other cases a large anamorphic factor could cause undersampling 
of the PSF (instrumental and seeing) on the detector. 
Ideally, the change in sampling caused by an increase in beam size could be 
matched to the decrease in image resolution. 
A decrease in beam size cannot cause undersampling on the detector. 

With a limited number of gratings, it would be useful to be able 
to use either a transmission or a reflection grating setup, with an increase 
or decrease in beam size. In this way, the aperture width 
could be matched more easily to the desired spectral resolving power. 

A decrease or no change in beam size after dispersion has the advantage 
of instrumental compactness, in that, the camera optics could be smaller 
than if an increase in beam size is allowed. 

\subsection{Finesse}

Choosing the finesse is tradeoff between a number of factors. 
A lower finesse has the advantage of needing less 
scans to complete the data cube. In effect, the total throughput 
of the instrument is approximately proportional to $1/N$. 
A higher finesse has the advantage of higher spectral and image resolution, 
and better suppression of the light between the narrow bands. 

\subsection{Comparison with a mono-order FP} 

The advantages of the TEI over a FP, where a single order has been 
selected with a blocking filter, include: 
imaging in many narrow bands simultaneously, and; 
imaging at any wavelength and spectral resolution without the
need for many order-sorter filters.
The disadvantages include:
a smaller field of view;
compromised image resolution (in one direction) because
of the PSF caused by the dispersion, and;
efficiency loss for a single order due to the extra
dispersive elements.

A significant aspect of the TEI is that the field, although smaller, 
will be monochromatic or nearly monochromatic in many cases. 
In fact, the size of the field decreases as the spectral resolving 
power ($R$) is increased which is in the same sense as the 
\citeauthor{Jac54} (\citeyear{Jac54}, \citeyear{Jac60}) criterion 
for a monochromatic field:
\begin{equation}
R \, \Omega = 2 \pi \: ,
\end{equation}
where $\Omega$ is the solid angle at the FP. 
The TEI parameters can be chosen to give maximum spectral coverage 
for the monochromatic field. 

\subsection{Comparison with the GraF concept}

The GraF concept uses a single dispersive element to separate the 
FP orders \citep{CL95}. 
Such a concept has been used in the GRAF/ADONIS instrument 
on the ESO 3.6-m telescope 
(\citeauthor{CLL99} \citeyear{CLL99}, \citeyear{CLR99}). 
This instrument operates in the near-IR with adaptive optics used 
to obtain 0.1\,arcsec image resolution. 
As an example, it can provide 10 images of 9'' by 0.9'' in the $H$ band 
at a spectral resolution of 10000 \citep{CLR99}. 
The field is thus small and narrow for high image resolution. 

Here, we compare the TEI with a more general GraF instrument.
A lower image resolution GraF instrument could, for example, have 
a field of 6\,arcmin by 20\,arcsec. 

The advantage of the TEI is the 
ability to image in more FP orders simultaneously,
i.e., providing a larger wavelength coverage except for the lowest spectral
resolutions.
The disadvantages include:
efficiency losses due to the extra dispersive element, and;
a shorter field of view.

In general, the shorter field of view is not a significant disadvantage 
because (i) the field will not be monochromatic over a longer field 
of view and (ii) astrophysical objects of interest may not be extended 
in one direction only, over the length of field possible 
with a GraF instrument (arcminutes). 

The TEI could be converted to a GraF instrument by replacing the echelle 
grating with a mirror, or positioning the camera for direct imaging. 

\subsection{Comparison with an echelle grating}

The advantages of the TEI over a standard echelle grating include:
imaging, and;
better spectrophotometry because a wider aperture can be used
without compromising wavelength calibration or resolution.
A disadvantage is that it 
requires scanning (tens of exposures) to fully sample a wavelength region.

The TEI can reach the same resolution as a standard echelle with a wider 
aperture and a smaller beam. The first factor allows better 
sampling of extended sources and the second allows the TEI to be a 
Cassegrain instrument with a smaller camera. 
These two factors partially offset the disadvantage of the scanning 
requirement in terms of efficiency. 
In certain cases, the TEI could be comparable to a slit echelle 
spectrograph for high-resolution spectroscopy of extended sources. 
For spectrophotometry and imaging, the TEI is significantly better.
 
\subsection{Comparison with an IFS}

In recent years, there has been great interest in the use of
integral-field spectrographs (IFS).  These include image slicers
\citep[e.g.,][]{Con00}, integral-field units and the TIGER approach
\citep{Cour82,CGM88,BAB95}.  Both TEI and IFS instruments are true 3D
spectrographs. However, they differ in one key respect. For the IFS,
the incident wavefront is divided and dispersed, in contrast to the
TEI where the wavefront is only dispersed.

The advantages of the TEI compared to an IFS include: 
better imaging and spectrophotometry because images do not need 
to be reconstructed;
a larger field of view for a given detector size, and;
potentially higher spectral resolution.
The disadvantages include:
the scanning requirement to fully sample a wavelength region, and;
compromised image resolution (in one direction) because of the PSF.

Integral-field spectrographs divide the focal plane and therefore 
image integrity can be compromised (i.e., they are not spectrophotometric).
So while an IFU is more efficient than a TEI device, 
for the same image size and spectral resolution requirements, 
there are certain cases where the TEI's undivided approach to 
3D spectro-imaging could be more beneficial. 

\subsection{Some applications}

The potential applications include: 
spectro-imaging and/or high-resolution spectroscopy of extended sources, 
through production of a full data cube; 
OH suppression in the 600--1600\,nm wavelength range ($RIzJH$), and;
targeted line imaging.

\begin{figure}
 \plotone{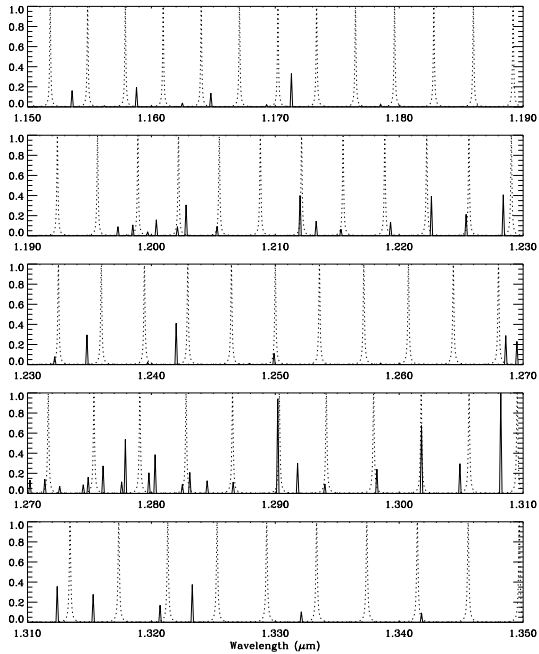}
 \caption{Plots of the transmission profile of a FP (dotted line)
   with a gap spacing of 220\micr\ and a finesse of 30, and an
   OH sky emission spectrum (solid line), in the $J$ band.
   The TEI images produced by the different FP orders can be combined
   selectively to exclude most of the OH emission.  The sky emission
   spectrum is the same as that used by \cite{OB98}.}
 \label{fig:oh-suppress}
\end{figure}

The OH-suppression method is to selectively combine the images produced 
by a TEI, rejecting those images that contain OH-line emission 
(or applying a lower weight). 
This will considerably reduce the sky background especially in 
the $J$ and $H$ bands. 
Figure~\ref{fig:oh-suppress} shows an example of a FP transmission 
profile plotted over the $J$-band sky emission spectrum. 

Evaluation of OH-suppression methods has been done by
\cite{Her94,JBB96,OB98}.  Equation~5 of \citeauthor{JBB96} gives the
gain ($G$) in performance of an OH-suppression filter for the
broad-band S/N.\footnote{The analysis of \cite{JBB96} uses
  Poisson-noise statistics. They point out that even $G \sim 1$ could
  provide a significant S/N gain, if most of the OH emission is
  removed from the background, because the continuum background is
  more stable and therefore systematic effects could be reduced.} 
For background limited conditions, this equation reduces to
\begin{equation}
 G = \frac{\tau}{\sqrt{\beta}} \: ,
\end{equation}
where $\tau$ is the fraction of object flux remaining, and $\beta$ is the 
fraction of background remaining. 
Simple numerical calculations, for the broad-band S/N in the $J$ band, 
show that $G$ can be larger or about unity, when selectively combining 
TEI images at resolving powers of above 10000 using a finesse of 30. 
Note that this broad-band gain only applies to dark time because the 
TEI is imaging between the OH lines where the background flux is 
moon-phase dependent. 
Additionally, a suitable TEI instrument could image in the 
$J$ and $H$ bands simultaneously and provide a spectrum 
(low-resolution if the FP is not scanned).

As with a mono-order FP, the TEI could be used for targeted line imaging 
but, in addition, the TEI can simultaneously provide an OH-suppressed 
broad-band image using the non-targeted FP orders. 
The reduction of field of view in comparison with a standard FP mode 
is not always significant because, if the FP is not scanned, it is only 
the monochromatic field of view that images the targeted emission 
line. 

\section{Summary}

The Tunable Echelle Imager offers an interesting reformation of existing 
spectrographs, by the use of a Fabry-Perot filter crossed with an 
echelle grating. 
This forms a mosaic of narrow-band images on a detector, offering a 
significant multiplex wavelength advantage over a mono-order FP system. 

The TEI instrument is placed in the collimated beam and includes 
a FP element, a grism and an echelle grating. 
Parameters such as the plate separation of the FP and the dispersion 
of the echelle determine the resolving power and the field-of-view size 
of the images. 
Resolving powers ranging from about 1\,000 to 100\,000 are observable 
with fields of view ranging from an arcminute down to arcseconds in size. 

The potential of the TEI includes its spectrophotometric integrity 
for producing a 3D datacube and the ability to create 
an OH-suppressed image by selectively combining different FP orders. 

\section*{Acknowledgments}

We would like to thank Keith Taylor and Brian Boyle for discussions.

\end{document}